\documentclass[12pt]{article}
\usepackage{graphicx}
\input german
\begin{document}

\centerline {\bf \large Computer-Simulation des Wettbewerbs zwischen Sprachen}

\bigskip

\centerline{D. Stauffer$^1$, C. Schulze$^1$, S. Wichmann$^2$}

\centerline{$^1$ Theoretische Physik, Universit"at, D-50923 K\"oln, 
Euroland}

\centerline{$^2$ Abteilung f"ur Linguistik, Max Planck Institut f"ur 
Evolution"are Anthropologie}
\centerline{Deutscher Platz 6, D-04103 Leipzig, Germany}

\bigskip

\noindent
{\bf Zusammenfassung}

\noindent
Recent computer simulations of the competition between thousands of languages
are reviewed, and some new results on language families and language 
similarities are presented. 

\bigskip
\noindent
{\bf 1. Einleitung}

\noindent
Bereits im Jahre 2006 hat K"olner Universit"at die Integration der 
passenden Didaktiken in die Mathematisch-Naturwissenschaftliche Fakult"at 
vollzogen, was im Saarland zu der Zeit eingeleitet wurde, als DS dort war
(1974-77, Gruppe Binder). Deshalb und weil DS auch jahrzehntelang die 
Lehramtskommission der K"olner Fachgruppe Physik leitete und das Lehramtsstudium
gegen die diversen Vorgaben von oben verteidigte, sind wir jetzt berechtigt, 
Herrn Patt zum 70. Geburtstag zu ehren.  

\begin{figure}[hbt]
\begin{center}
\includegraphics[angle=-90,scale=0.30]{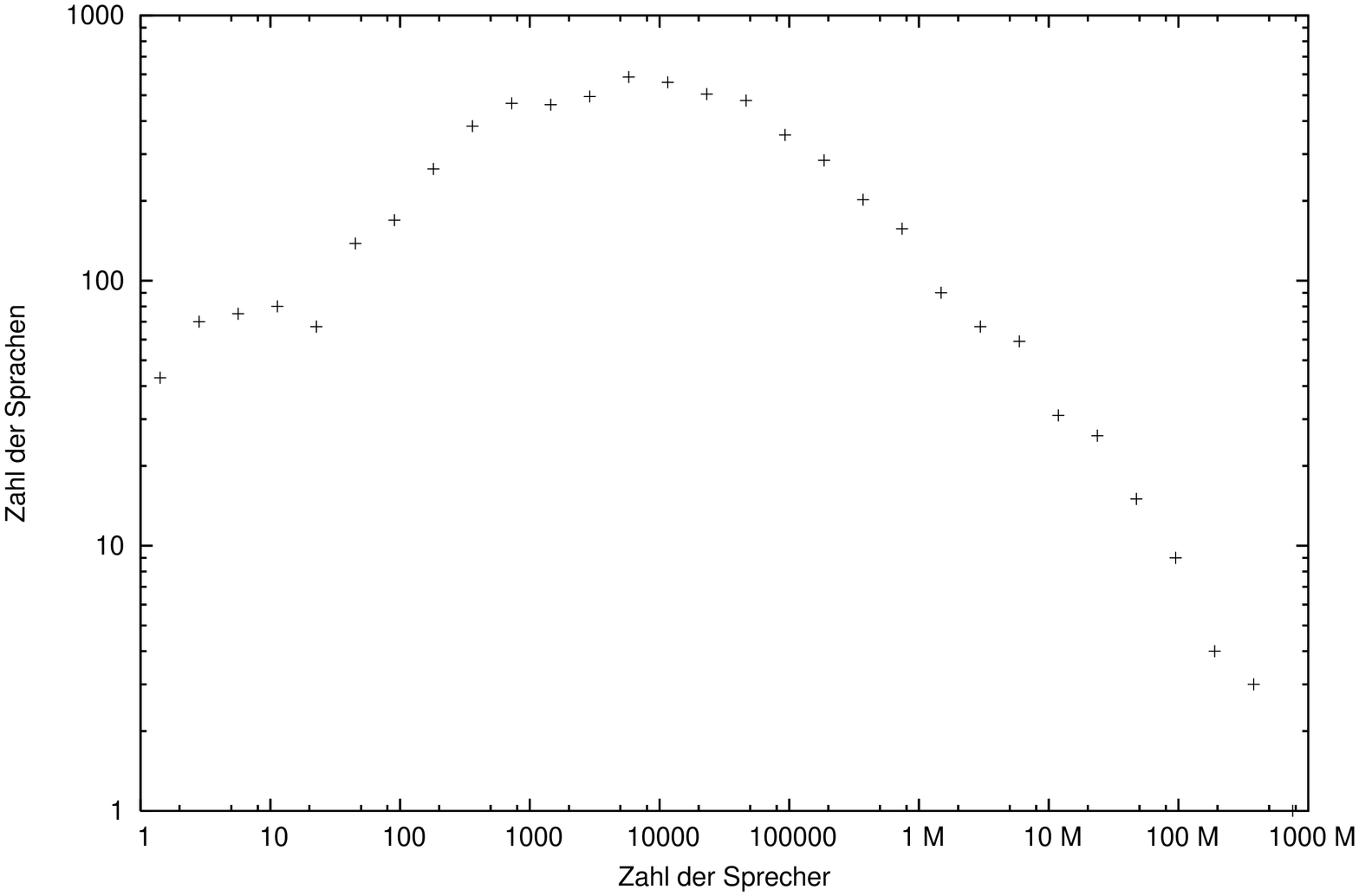}
\includegraphics[angle=-90,scale=0.30]{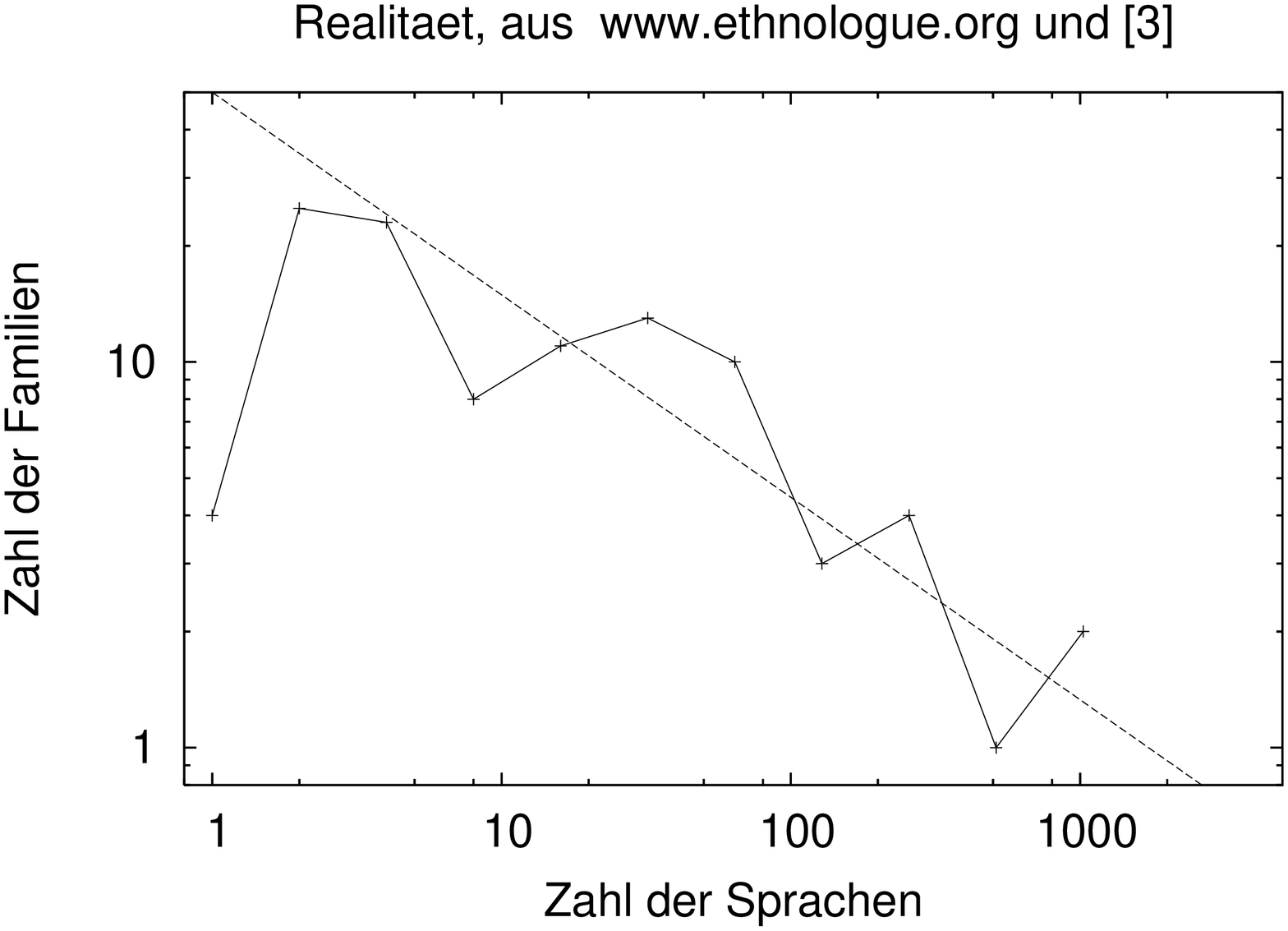}
\end{center}
\caption{Empirische Verteilung der realen Sprachgr"o"sen \cite{chak} (oben) 
und Familiengr"o"sen \cite{wichmann} (unten). }
\end{figure}

Theoretische Physik ist heute auch nicht mehr das, was sie fr"uher einmal war. 
Auf dem Computer sind bis zu $10^{13}$ Teilchen simuliert werden, und so
k"onnen komplexe System simuliert werden, wo das Verhalten des Gesamtsystems
nicht direkt aus dem Verhalten der einzelnen Teilchen erschlossen werden kann.
Ein Beispiel daf"ur sind reale Gase, die seit 1872 durch die Gleichung von
van der Waals approximiert werden. Aus der Tatsache, dass sich die Teilchen 
teilweise anziehen und teilweise absto"sen, kann man nicht direkt erkennen,
dass es ein Gleichgewicht von Dampf und Fl"ussigkeit und eine kritische 
Temperatur gibt. Auch der Unterschied von Massenpsychologie zur Psychologie
der Einzelnen kann als Komplexit"at gelten. 

So haben in den letzen Jahren die Physiker auf ihren Computern nicht nur 
Biologie simuliert (Biologie ist eine alte Liebe der Physik, die erst seit
kurzem erwidert wird), sondern auch B"orsenkurse, Meinungsbildung, soziale
Netzwerke, $\dots$ \cite{newbook}. Seit 2004 hat diese Seuche auch die 
Sprachwissenschaft erreicht \cite{abrams}, und umgekehrt benutzen Linguisten 
auch Methoden aus der Physik komplexer Systeme \cite{wichmann}. Diese Arbeiten 
\cite{abrams} f"uhrten kurz danach an sechs verschiedenen Unis unabh"angig von 
einander zu weiteren Simulationen des Wettbewerbs zwischen Sprachen.
"Uber diese Computersimulation der Wechselwirkung zwischen verschiedenen
Sprachen wird hier berichtet; eine ausf"uhrlichere aber schon veraltende 
"Ubersicht ist \cite{chak}. Wir benutzen die seit einem halben Jahrhundert 
in der Physik vorhandenen Methoden der Simulation einzelner Teilchen, ohne
dass wir n"aherungsweise "uber viele Teilchen mitteln; in anderen 
Wissenschaften ist diese Methode neuer und wird oft ``Agenten-basiert''
genannt \cite{fent}.  "Ubrigens hat Selten, Wirtschaftsnobelpreistr"ager in
Bonn, schon viel fr"uher Spieltheorie auf Sprachen angewandt \cite{selten}.

Heute sprechen die Menschen etwa 7000 Sprachen, davon alleine etwa 800 in
Papua-Neuguinea. Alle zehn Tage stirbt im Durchschnitt eine Sprache aus.
Werden in einigen Jahrtausenden alle Menschen die gleiche Sprache (und ihre 
Varianten) sprechen \cite{parijs}? 
K"onnen wir die reale Verteilung der Sprachgr"o"sen in 
einer Simulation reproduzieren: Wieviel Sprachen gibt es, die jeweils $N$
Leute als Muttersprache haben? Diese reale Verteilung in Abb.1a ist eine
Log-Normal-Verteilung mit einer Erh"ohung bei den ganz kleinen Sprachen, 
die nur noch von ein paar Leuten gesprochen werden. Eine 
Parabel in dieser doppelt-logarithmischen Abbildung entspricht einer 
lognormalen Verteilung. Mandarin-Chinesisch wird von mehr als einer Milliarde
gesprochen, und etwa 50 Sprachen nur noch von einer Person. 

Im n"achsten Abschnitt 
werden die beiden Modelle erkl"art, mit denen wir 7000 Sprachen gleichzeitig 
simulieren k"onnen, und im "ubern"achsten Abschnitt bringen wir ausgew"ahlte 
Resultate. "Uberall z"ahlt die Gr"o"se einer Sprache die Zahl der Sprecher 
dieser Sprache. Ein Anhang listet eines der benutzten Fortran-Programme.

\bigskip
\noindent
{\bf 2. Vielsprach-Modelle} 

\noindent
2.1 Schulze-Modell

\noindent
Dieses erste Modell f"ur viele Sprachen gibt es in diversen Versionen, deren
Resultate meist "ahnlich sind; hier wird die letzte Version erkl"art. Auf jedem
Gitterpunkt eines $L \times L$ Quadratgitters lebt ein Mensch, der genau
eine Sprache spricht. Diese Sprache wird definiert durch $F$ Eigenschaften,
von denen jede durch eine ganze Zahl zwischen 1 und $Q$ characterisiert
ist. Beispiele sind grammatische Eigenschaften wie die Reihenfolge von Subjekt, 
Objekt und Verb. Typische Werte sind $Q = 2$ bis 5, $F = 8$ oder 16. Bei $Q=2$ 
kann man effizienter rechnen, wenn $F = 8, 16, 32$ oder 64 Eigenschaften dann 
in einem einzigen Computerwort (Bitstring) als Bits abgespeichert werden. Bei 
einer Iteration wird jeder Gitterpunkt einmal behandelt.
 
Mit Wahrscheinlichkeit $p$, unabh"angig f"ur jede der $F$ Eigenschaften, 
"andert sich die betrachtete 
Eigenschaft der am betrachteten Gitterpunkt gesprochen Sprache. 
Diese Wahrscheinlichkeit $p$ ist in der Realit"at etwa ein Prozent 
pro menschlicher Generation.
Au"serdem springt mit Wahrscheinlicheit $1-x^2$ oder $(1-x)^2$ jemand von einer
Sprache, die von einem Bruchteil $x$ der Gesamtbev"olkerung gesprochen wird, 
zur Sprache einer zuf"allig ausgew"ahlten (benachbarten) Person. Letzterer 
Prozess ist eine typisch menschliche Eigenschaft, dass man von einer 
``kleinen'' zu einer weit verbreiteten Sprache springt, wie es in der 
Physikforschung in der zweiten H"alfte des 20. Jahrunderts geschah. 

Dar"uber hinaus k"onnen auch (mit Wahrscheinlichkeit $q$) Eigenschaften von 
einem zuf"allig ausgew"ahlten Nachbarn "ubernommen werden. Beispiele 
f"ur diese linguistische Diffusion sind franz\-"osische Lehnworte oder englische
Grammatik im Deutschen: ``Ich ging in 2005 zweimal zum Friseur''. Und obiger
Sprung von einer zur andereren Sprache wird nur mit Wahrscheinlichkeit $r$
bedacht, also insgesamt mit Wahrscheinlichkeit $(1-x)^2r$ gemacht. In Abschnitt
3.1 werden wir noch L"ocher ins Schulze-Modell bohren oder es auf ein 
Barab\'asi-Albert Netzwerk setzen.

\bigskip
\noindent
2.2 Viviane Model

Die Kolonisierung eines zun"achst menschenleeren Kontinents wurde von Viviane 
de Oliveira et al modelliert \cite{viviane} und wird hier brasilianischem Stil 
entsprechend mit dem Namen Viviane bezeichnet. Zun"achst beschreiben wir
die urspr"ungliche Version \cite{viviane}, dann eine bessere Modifikation in
zwei Varianten a und b \cite{pmco}. 

\bigskip
\noindent
2.2.1 Urspr"ungliche Version

\noindent
Auf jedem Gitterpunkt $j$ eines $L \times L$ Quadratgitters leben entweder $c_j$
Menschen, oder niemand, mit $1 \le c_j \le m$ zuf"allig gew"ahlt, und $m \sim 
10^2$. Auf einem bewohnten Gitterplatz wird nur eine Sprache gesprochen. 
Anfangs ist nur ein Gitterpunkt bewohnt, so wie Amerika wohl von der 
Beringstra"se aus besiedelt wurde. Danach breitet sich die Bev"olkerung aus, 
Schritt f"ur Schritt in einen freien Nachbarn bewohnter Pl"atze; dieser wird mit
einer Wahrscheinlichkeit $c_j/m$ ausgew"ahlt. Dort wird erst die Sprache $k$
eines der bewohnten Nachbarpl"atze gesprochen, der mit einer Wahrscheinlichkeit 
proportional zur Sprachgr"o"se $N_k$ (``Fitness'') ausgew"ahlt wird. 
Anschlie"send "andert sich mit einer Wahrscheinlichkeit $\alpha/N'_k$
die Sprache am neu besiedelten Platz, so dass dort eine neue entsteht, 
wobei $N'_k$ die kleinere der beiden Zahlen $N_k$ und $M_k$ ist, und die 
Grenze $M_k$ anfangs zuf"allig zwischen 1 und $M_{\max} \sim 20 m$ festgelegt
wird. (In anderen Worten: Die Mutationsrate ist umgekehrt proportional zur 
Sprachgr"o"se, mindestens aber eine von der Sprachgr"o"se unabh"angige und
zuf"allig am Anfang festgelegte Wahrscheinlichkeit $\alpha/M_k$.) 
Der Mutationsfaktor $\alpha$ ist ein freier Parameter. Die neue Sprache 
bekommt eine neue Nummer; andere mehr inhaltliche Sprachelemente hat das 
urspr"ungliche Modell nicht. Die Sprachen auf den bereits besetzten Pl"atzen 
"andern sich nicht mehr. Die Simulation endet, wenn alle Gitterpl"atze besetzt 
sind.

\bigskip
\noindent
2.2.2 Modifizierte Versionen

Obiges Viviane-Modell wird beibehalten bis auf folgende zwei F"alle von 
Modifikationen, die zu zwei Versionen f"uhren und auf Paulo Murilo de 
Oliveira \cite{pmco} zur"uckgehen:

\noindent
a) Zun"achst kann jeder Sprache des Viviane-Modells ein Bitstring zugeordnet 
werden, also eine Kette von 8 bis 64 bin"aren Variablen in den bisherigen
Simulationen (siehe Programm im Anhang). Diese Bits geben der Sprache einen 
Inhalt und erlauben, die Unterschiedlichkeit verschiedener Sprachen zu 
bestimmen (siehe Abschnitt 3.2). Wie zuvor f"uhrt jede Mutation bei der
Besiedlung eines neuen Platzes zu einer neuen Sprache.

\noindent
b) Stattdessen kann auch eine Sprache nur dann als neu definiert werden, wenn 
der Bitstring sich "andert und eine bisher nicht aufgetretene Folge von Bits
darstellt. Die Zahl der verschiedenen Sprachen ist dann die Zahl der 
verschiedenen Bitstrings. Die "Anderung einer Sprache kann damit auch zu 
einer bereits vorhandenen Sprache f"uhren, etwa wenn das erste Bit von 01011000
von Null auf Eins springt und der Bitstring 11011000 schon auf einem anderen 
Gitterplatz mit einer anderen Vorgeschichte realisiert ist. Au"serdem wird
ber"ucksichtigt, dass die meisten Gegenden der Erde weniger attraktiv sind als
die gro"sen Metropolen; die Bev"olkerungsdichte $c$ zwischen 1 und $m$ wird 
nicht gleichf"ormig bestimmt, sondern mit einer Wahrscheinlichkeit proportional
zu $1/c$, analog zu realen Gr"o"senverteilung von St"adten. Dann ist es 
effizienter, die Auswahl eines unbewohnten Nachbarplatzes proportional zu seiner
Attraktivit"at $c$ dadurch zu realisieren, dass zwei solche unbewohnte Nachbarn
zuf"allig ausgew"ahlt werden, und der mit dem gr"o"seren $c$ anschlie"send
besiedelt wird.

\begin{figure}[hbt]
\begin{center}
\includegraphics[angle=-90,scale=0.32]{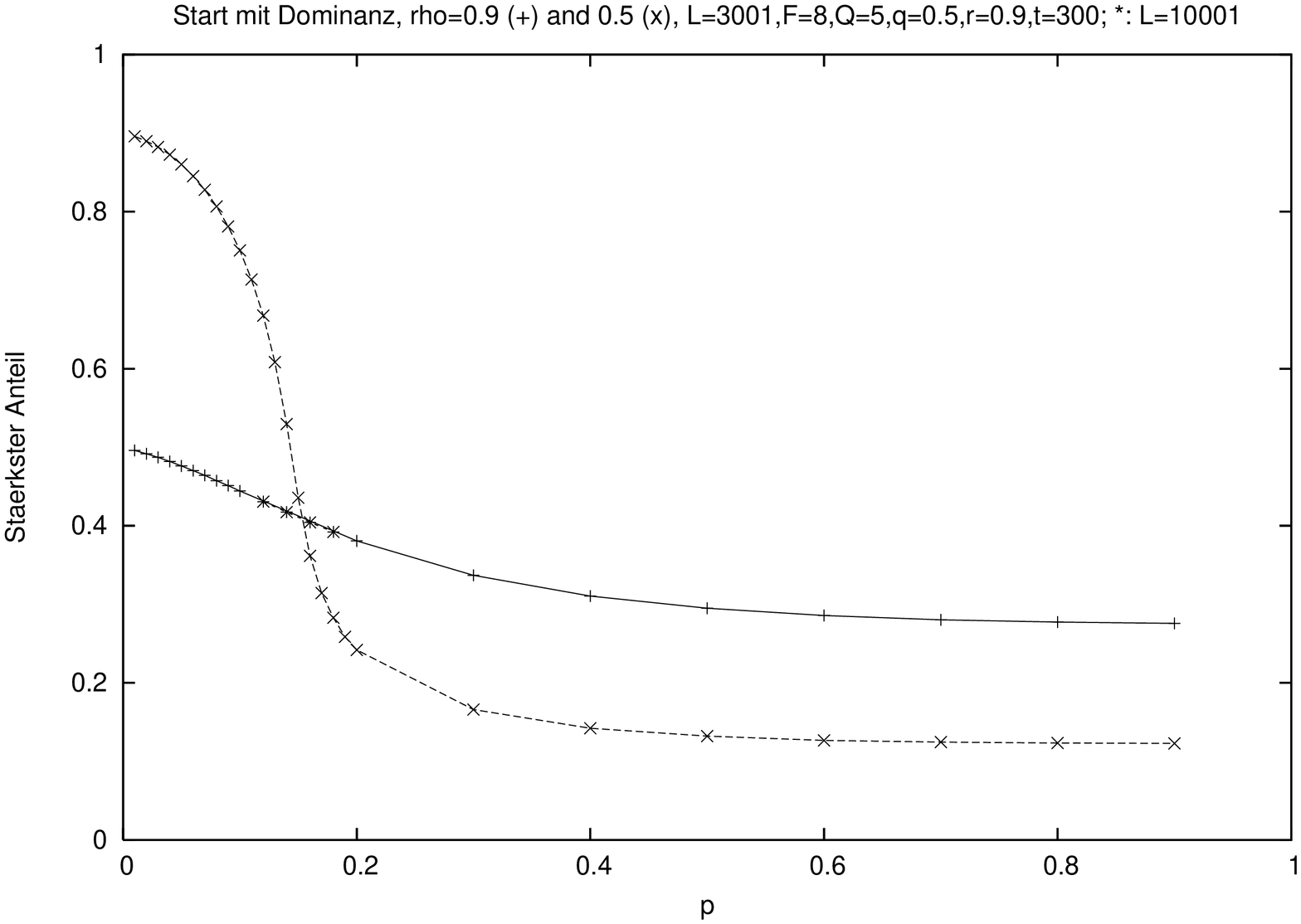}
\includegraphics[angle=-90,scale=0.32]{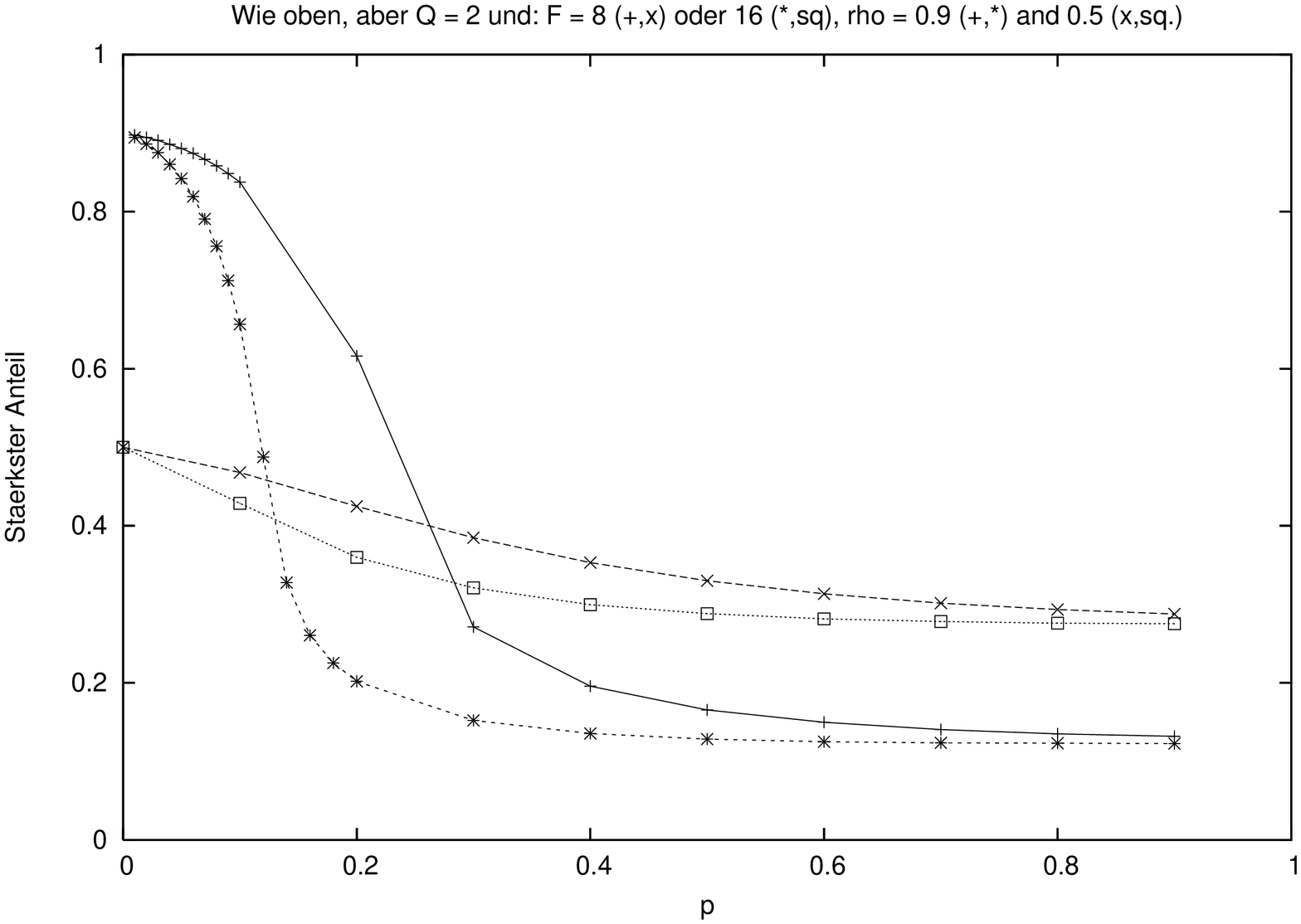}
\end{center}
\caption{Anteil der am Ende gr"o"sten Sprache an der Gesamtbev"olkerung im   
l"ochrigen Schulze-Modell.}
\end{figure}

\begin{figure}[hbt]
\begin{center}
\includegraphics[angle=-90,scale=0.29]{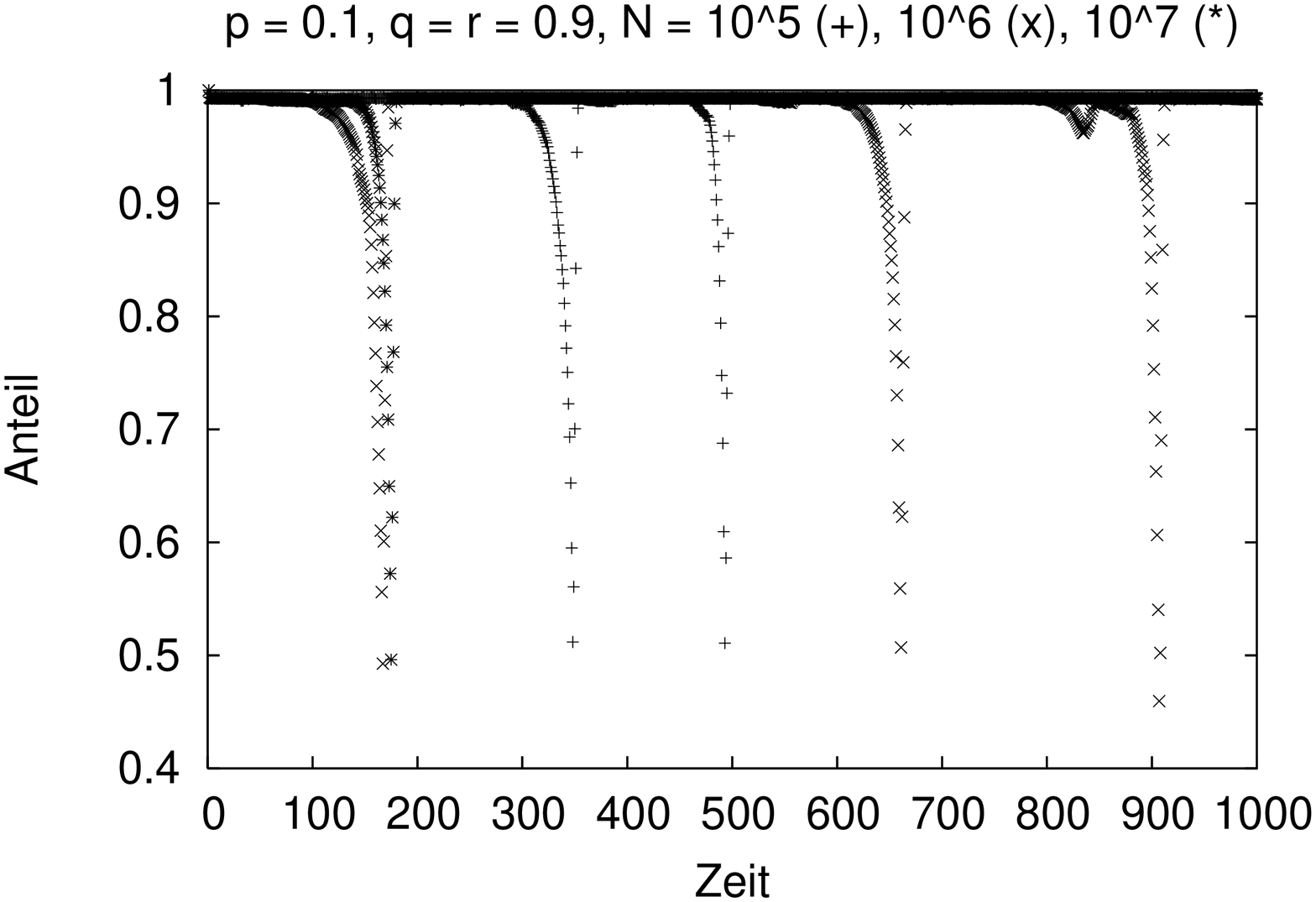}
\includegraphics[angle=-90,scale=0.29]{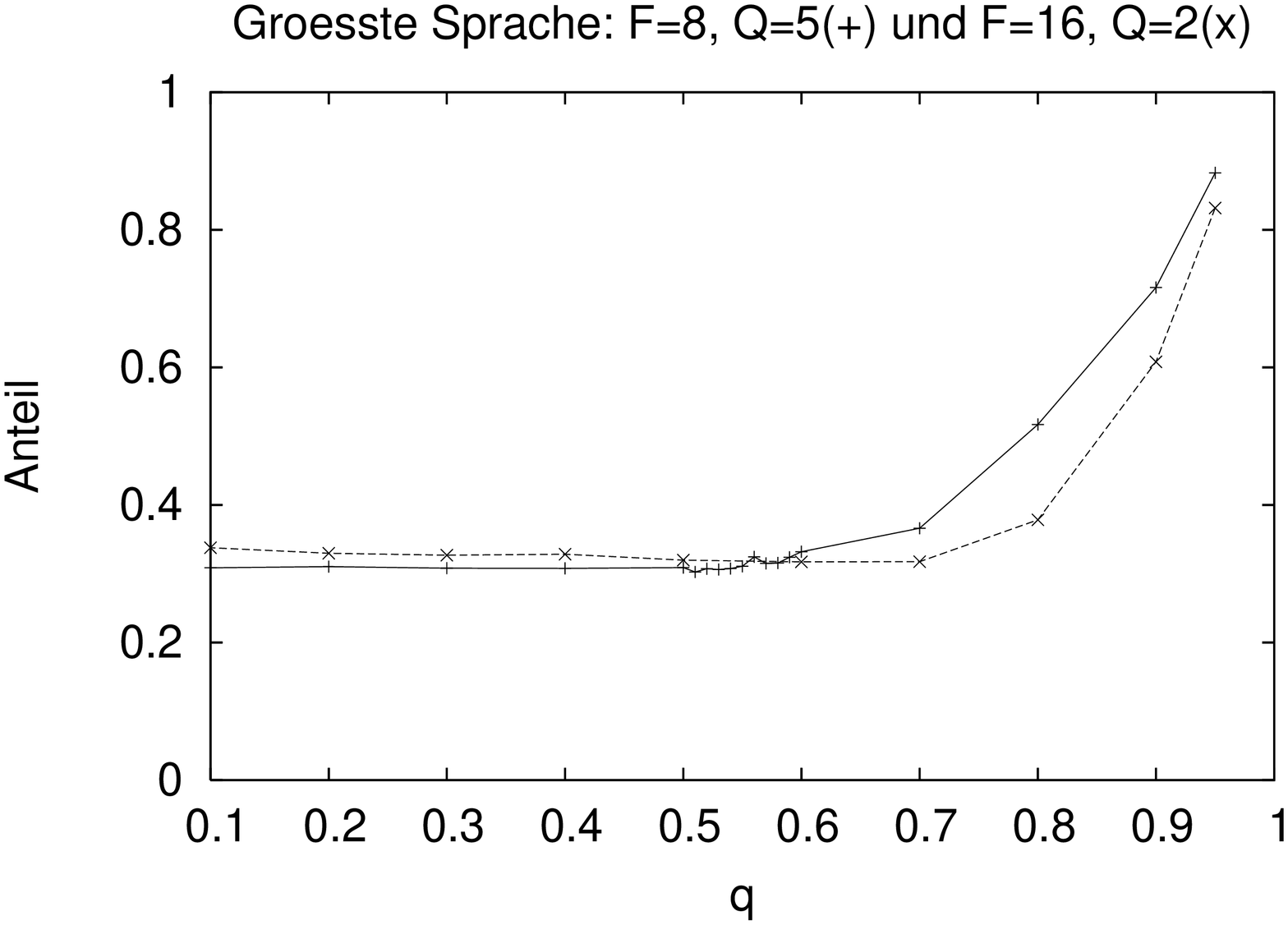}
\includegraphics[angle=-90,scale=0.29]{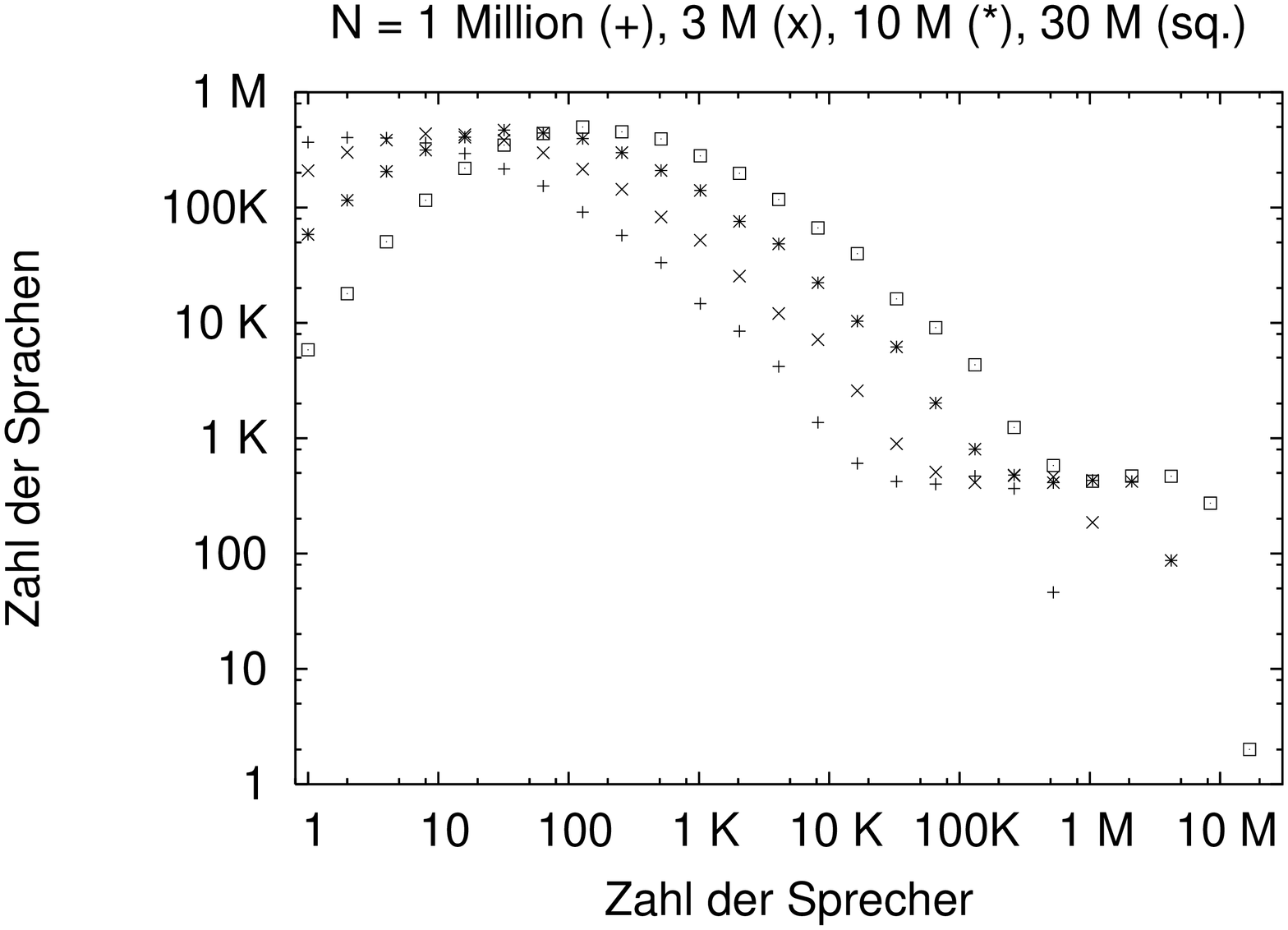}
\includegraphics[angle=-90,scale=0.29]{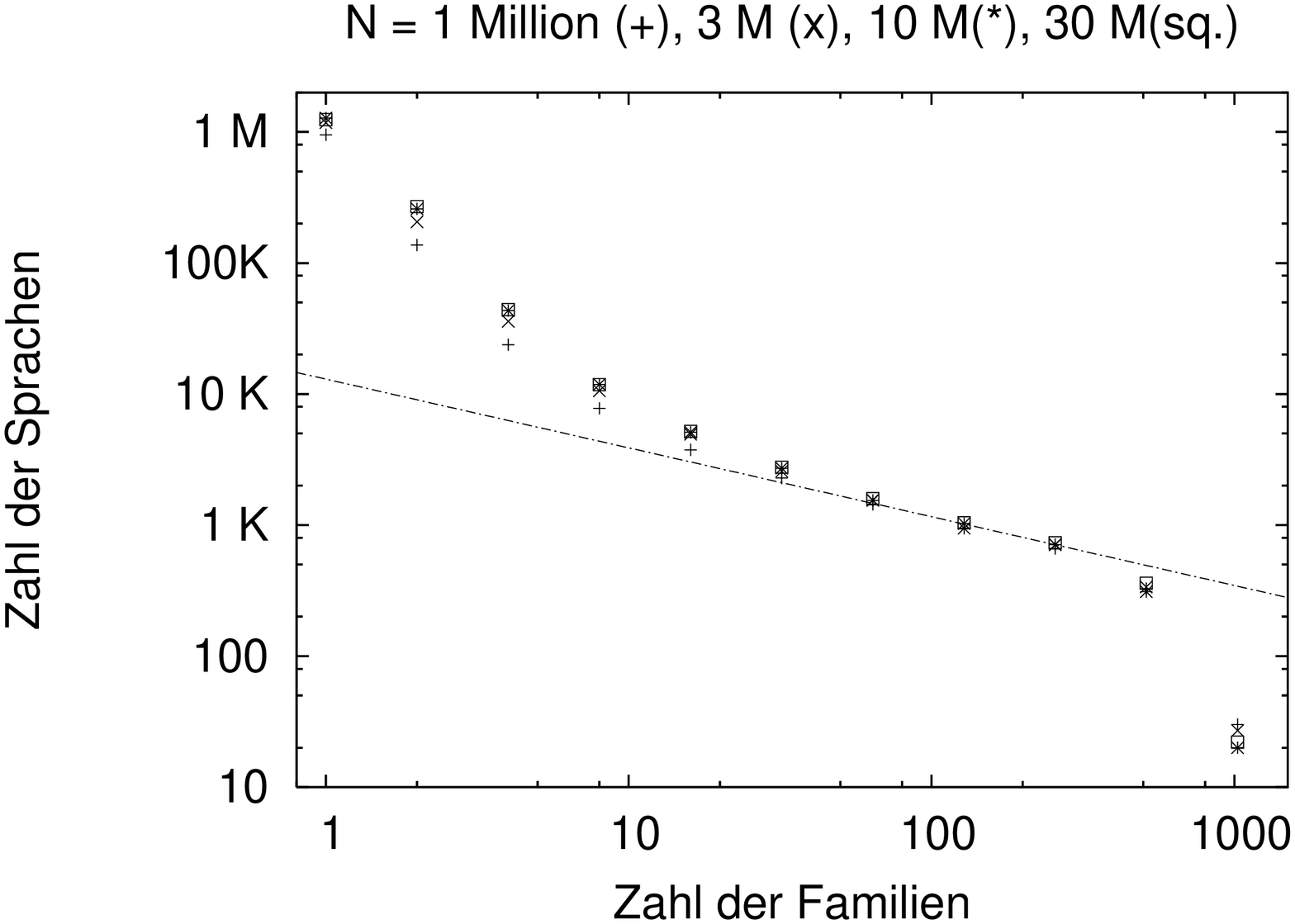}
\end{center}
\caption{$F=8, \; Q=5, \; 10$ Millionen Leute: Anteil der gr"o"sten Sprache an
der Gesamtbev"olkerung (oben) und Gr"o"senverteilung (unten), im 
Schulze-Modell auf gerichtetem Barab\'asi-Albert Netwerk. Rechts und unten: 
$p=0.5, \, q=0.59, \, r=0.9$; Steigung --0,525.}
\end{figure}

\begin{figure}[hbt]
\begin{center}
\includegraphics[angle=-90,scale=0.5]{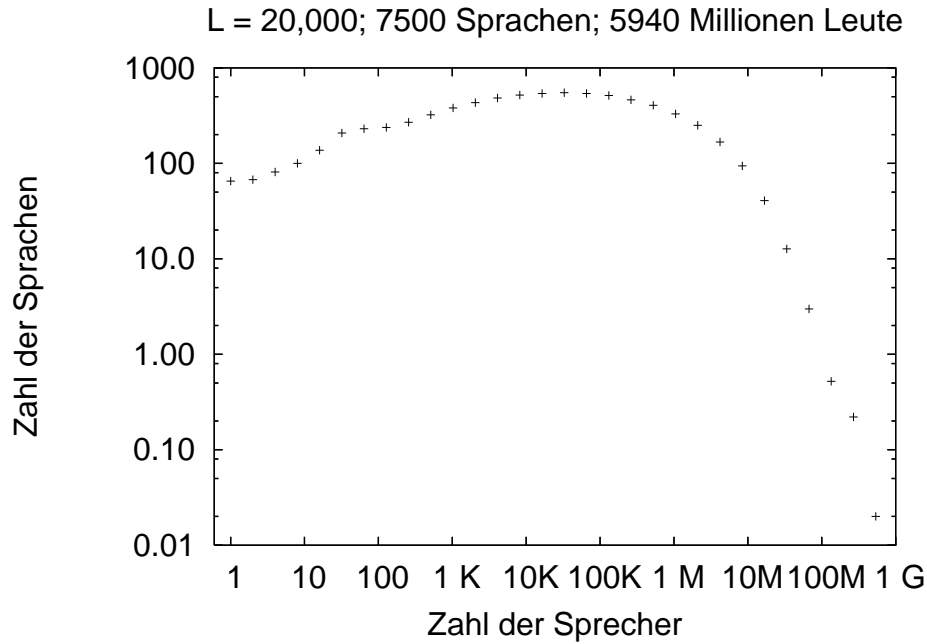}
\end{center}
\caption{Gr"o"senverteilung der Sprachen im modifizierten Viviane-Modell, Fall 
b, 13 Bits pro Bitstring, $\alpha = 0.1, \, m = 63, \, M_{\max} = 255$.}
\end{figure}

\begin{figure}[hbt]
\begin{center}
\includegraphics[angle=-90,scale=0.31]{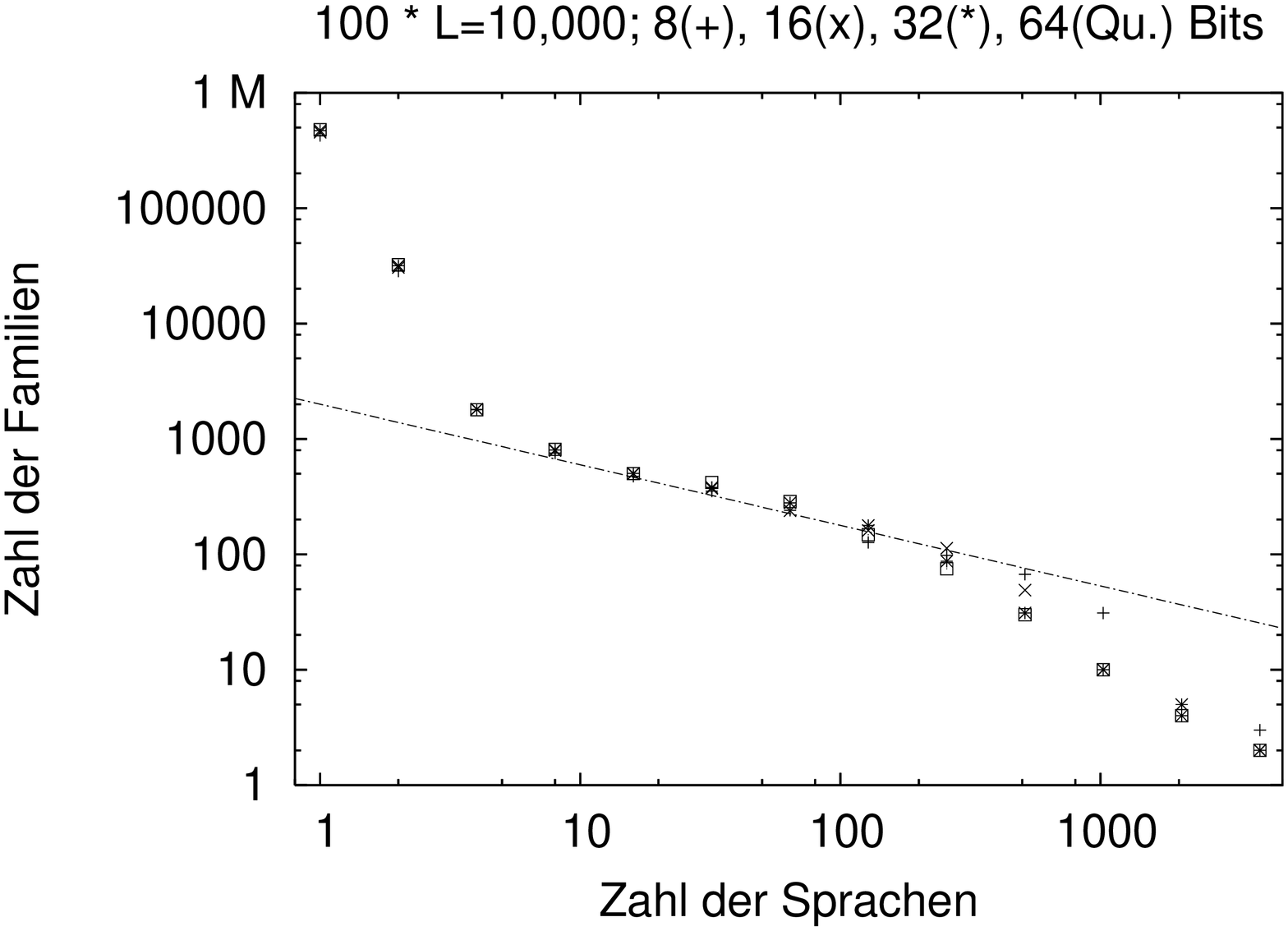}
\includegraphics[angle=-90,scale=0.31]{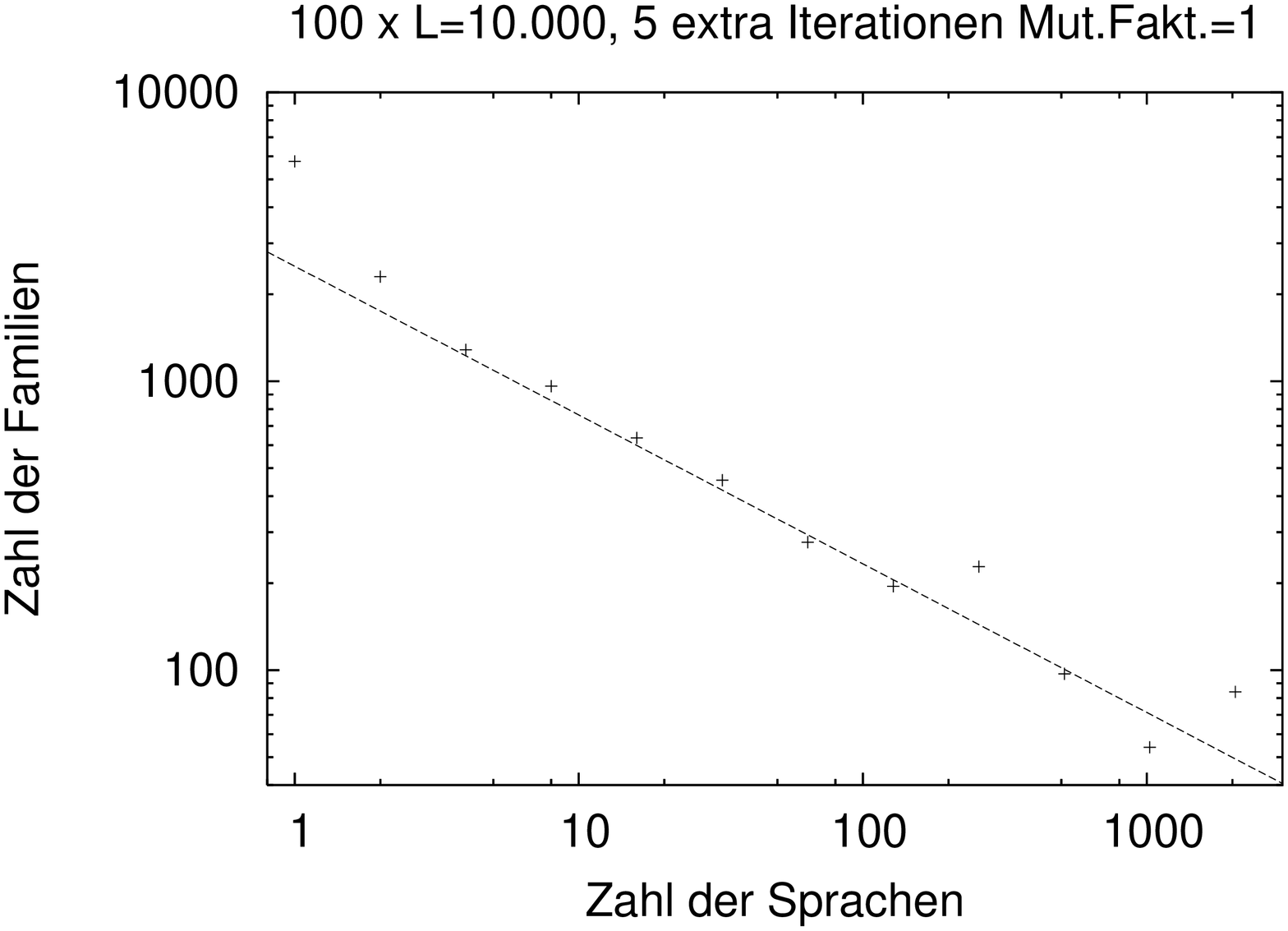}
\end{center}
\caption{Einteilung der Sprachen in Sprachfamilien im modifizierten 
Viviane-Modell, Summe "uber 100 Gitter, zu vergleichen mit Abb. 1b. 
Oben: Fall a,  $\alpha = 0.05, \, m = 127, \, M_{\max} = 255$. 
Unten: Fall b, $\alpha = 0.10, \, m =  63, \, M_{\max} = 255$. } 
\end{figure}

\begin{figure}[hbt]
\begin{center}
\includegraphics[angle=-90,scale=0.5]{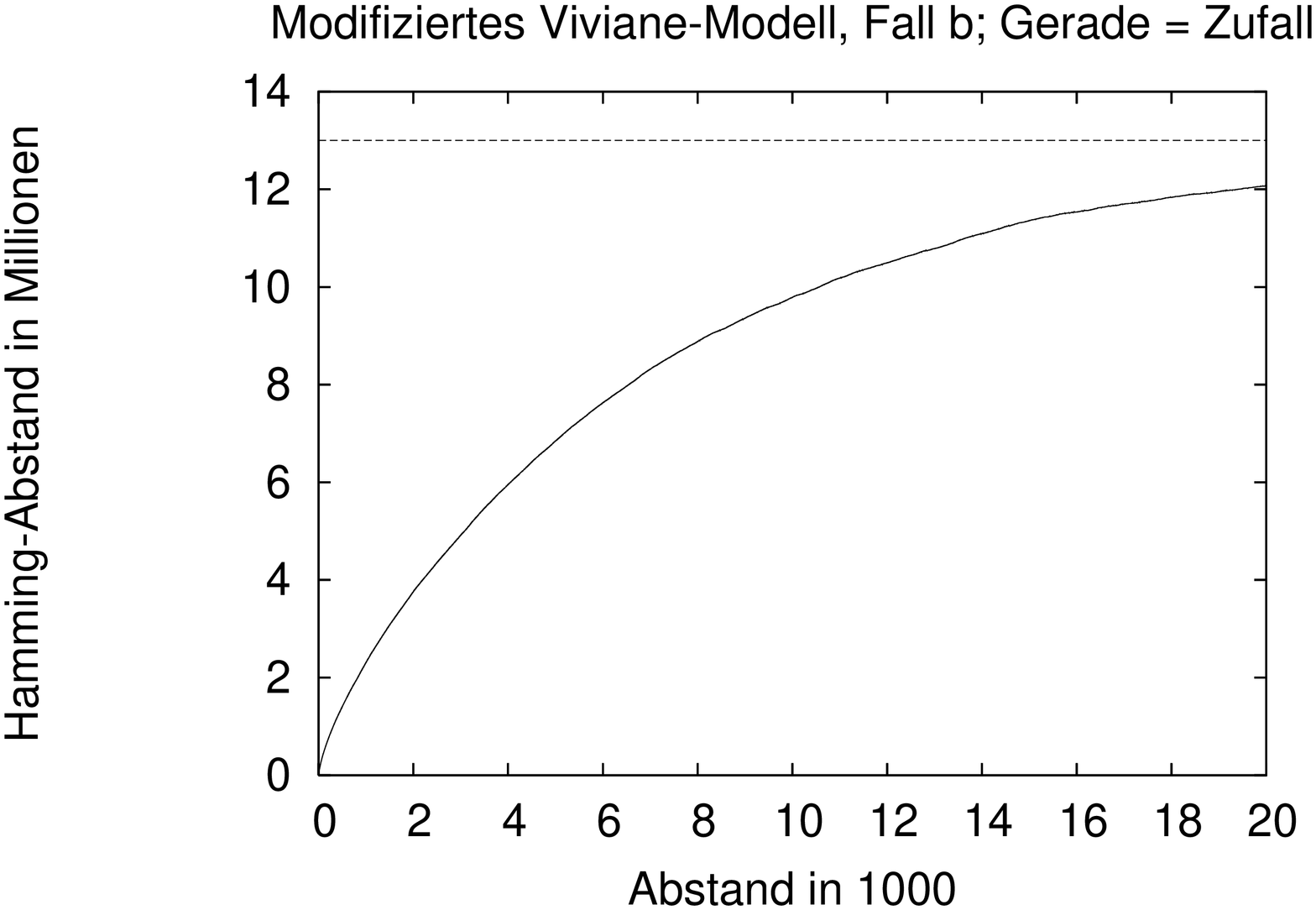}
\end{center}
\caption{Un"ahnlichkeit zwischen Sprachen als Funktion des geographischen 
Abstands. Die Einheit der horizontalen Achse sind 1000 Gitterkonstanten.
Die horizontale Gerade zeigt den summierten Hammingabstand an, wenn die 
Sprachen v"ollig unkorreliert w"aren. 13 Bits, $\alpha = 0.1, \, m = 63, \, 
M_{\max} = 255, \, L=20.000$.}
\end{figure}

\newpage
\noindent
{\bf 3. Resultate}

\noindent
3.1 Schulze-Modell

\noindent
Wenn anfangs Alle die gleiche Sprache sprechen, dann kann eine hinreichend hohe
"Anderungswahrscheinlichkeit $p$ zu einer Fragmentation f"uhren wie beim 
Turmbau zu Babel: Die Bev"olkerung verteilt sich etwa gleichm"a"sig auf die
$Q^F$ m"oglichen Sprachen. Man kann aber auch mit so einer fragmentierten 
Bev"olkerung beginnen und dann bei hinreichend kleinem $p$ sehen, wie nach 
einiger Zeit eine Sprache dominiert und von den meisten Leuten gesprochen wird;
die Anderen sprechen meist eine Variante dieser dominierenden Sprache. Wir 
sehen einen Phasen"ubergang erster Art, mit Hysterese.  Nur im 
Nichtgleichgewicht und mit zus"atzlichen Rausch-Effekten \cite{sol} konnte 
eine Gr"o"senverteilung "ahnlich zu Abb.1 simuliert werden.

Dinge wurden nicht besser, wenn Dreik"orper- und F"unfk"orper-Kr"afte 
angenommen wurden: Eine Eigenschaft wurde von einer anderen Sprache 
"ubernommen nur, wenn zwei, oder alle vier, Nachbarn die gleiche Eigenschaft 
hatten.

Ganz anders ist die Situation in der l"ochrigen Version, wenn nur ein 
zuf"alliger Anteil $\rho$ aller Pl"atze bewohnbar ist. 
``Bekanntlich'' bilden sich dann auf dem Quadratgitter
bei $\rho < 0,593$ nur endliche Cluster bewohnbarer Nachbarn, w"ahrend bei
$\rho > 0,593$ auch ein unendliches Cluster sich von einem Ende des Gitters zum 
anderen erstreckt. Abb.2 zeigt nun mit wachsendem $p$ einen kontinuierlichen 
Abfall des Anteils der (gegen Ende der Simulation von $t$ Iterationen) gr"o"sten
Sprache, "ahnlich zum Viviane-Modell: Die Unordnung hat den Phasen"ubergang 
zerst"ort. Die Gr"o"senverteilung der Sprachen stimmt aber immer noch nicht. 

Besser wird das Schulze-Modell, wenn statt auf zuf"allig besetzten
Pl"atzen die Sprecher auf ``scale-free'' Netzen vom Barab\'asi-Albert 
Typ sitzen \cite{ba}. Dabei geht man von drei miteinander 
verbundenen Netzwerk-Gr"undern aus, und addiert danach neue 
Mitglieder, eines nach dem anderen. Ein neues Mitglied sucht 
sich aus den zu diesem Zeitpunkt vorhandenen Mitgliedern
drei Vorgesetzte aus, proportional zur Zahl der Mitglieder,
die vorher diese Person als Vorgesetze gew"ahlt haben. 

Abb.3a zeigt jetzt statt einer dominierenden Sprache einen Wechsel 
von einer zu einer anderen dominierenden Sprache; w"ahrend des 
Wechsels sinkt der Anteil der gr"o"sten Sprache in einem kleinen Zeitinterval
auf etwa 1/2; dann wird eine andere Sprache dominierend, und ihr Anteil springt
wieder auf nahezu 1 hoch. Dies passiert etwa zweimal im betrachteten 
Zeitintervall $t = 1000$, f"ur jede der drei gezeigten Netzgr"o"sen in etwa 
gleicher Weise. In der Biologie nennt man das ``punctuated equilibrium''.
Wenn $p$ anw"achst, ergeben sich brauchbare Gr"o"senverteilungen
wie in Abb. 3c. (In Abb. 3 wird, wie am Ende von 2.1 erw"ahnt,
mit Wahrscheinlichkeit $q$ eine Sprach-Eigenschaft von einem 
Vorgesetzten "ubernommen, und der Sprung zu einer Sprache eines zuf"allig 
ausgew"ahlten Vorgesetzten geschieht mit Wahrscheinlichkeit $r(1-x)^2$.) 
Auch jetzt ist kein Phasen"ubergang mehr da. Die Gr"o"senverteilungen
f"ur die Sprachen (als Funktion der Zahl der Sprecher)  und f"ur die Zahl der 
Familien (als Funktion der Zahl der Sprachen) sind recht gut, Abb.3 unten.

\bigskip
\noindent
3.2 Viviane-Modell

\noindent
Im urspr"unglichen Viviane-Modell stimmt die Gr"o"senverteilung besser und 
erstreckt sich in gro"sen Gittern von 1 bis zu einer Milliarde. Nur gibt die
doppelt-logarithmi\-sche Darstellung keine Parabel, sondern zwei Geraden, die
zwei Potenzgesetzen entsprechen. Nur wenn wieder Rauschen \cite{sol} eingebaut 
wird, sieht die Verteilung ordentlich aus \cite{pmco}. 

Viel besser funktioniert die von Paulo Murilo de Oliveira (nicht mit Viviane
de Oliveira verwandt) modifizierte Version (Fall b in Abschnitt 2.2.2). Abb.4
aus \cite{pmco} zeigt eine leicht schiefe Parabel, "ahnlich zu Abb.1, mit 
6 Milliarden Menschen und 7500 Sprachen.   

Die Einteilung der Sprachen in Familien (z.B. die indogermanische Sprachfamilie)
funktioniert in beiden F"allen a und b. Eine neue Sprachfamilie startet genau 
dann, wenn
eine gerade mutierte Sprache sich um mindestens $i$ Bits von der historisch 
ersten Sprache dieser Sprachfamilie unterscheidet. Bei $i=1$ bildet jede neue 
Sprache eine neue Sprachfamilie: uninteressant. Von $i=2$, 3 und 4 funktioniert
$i=2$ am besten, und Abb. 5 zeigt oben $i=2$ mit Bitstrings der L"ange
8, 16, 32 und 64: Kein Einfluss der Zahl der Bits. Die Gerade hat die Steigung 
--0,525, die dem empirischen Exponenten von $-1,905 = -1/0,525$ der Realit"at
nach \cite{wichmann} entspricht. Der untere Teil zeigt Fall b mit 13 Bits,
wobei zu f"unf Zeiten im gleichen Abstand auch die l"angst besetzten Pl"atze
ihr Sprache "andern k"onnen, wie es dauernd bei den neu besetzten Pl"atzen 
geschieht.

Auch die Geographie spielt eine Rolle, und je weiter die Sprachen r"aumlich
voneinander getrennt sind, um so mehr unterscheiden sie sich im Durchschnitt.
Holman und Wichmann haben das f"ur die Realit"at untersucht \cite{holman},
und eine ganz "ahnliche Kurve liefert Fall b des modifizierten Viviane-Modells
in Abb. 6. Hier wird der Unterschied zwischen den Sprachen durch den 
Ham\-ming-Abstand zwischen den Bitstrings gemessen, also durch die Zahl der 
verschiedenen Bits bei einem Position-f"ur-Position Vergleich der beiden 
Bitstrings. Wenn die Sprachen gar nicht mehr korreliert w"aren, w"urden sie
in der H"alfte der Bits "ubereinstimmen, was durch die Gerade in Abb. 6 
symbolisiert wird. "Ahnlich zur Realit"at \cite{holman} sind die Sprachen 
erst dann nahezu unkorreliert, wenn wir von einem Ende des $20.000 \times 
20.000$ Gitters zum anderen gehen.
 
\bigskip
\noindent
{\bf 4. Diskussion}

\noindent
Nach vielen Anl"aufen sind im letzten Jahr sind erhebliche Fortschritte dabei 
gemacht worden, die
quantitative "Ubereinstimmung von Simulation und Realit"at zu verbessern. 
Die Zahl der Sprachen als Funktion der Zahl der Sprecher stimmt gut, Abb. 1 und
4, die der Zahl der Sprachfamilien als Funktion der Zahl der Sprachen in Abb. 1
und 5 stimmt einigerma"sen. Abb. 6 suggeriert, dass eine 
Gittereinheit knapp einem Kilometer entspricht, was auch mit der 
Bev"olkerungsdichte $1 \le c_j \le m \sim 10^2$ zusammenpasst. 
Man sollte nach alternativen Modellen suchen, die "Ahnliches leisten; 
derzeit ist das Schulze-Modell gegen"uber dem Viviane-Modell leicht
zur"uckgeblieben.

\bigskip
\noindent
{\bf 5. Anhang}

\noindent
Das folgende Fortran-Programm addiert zum urspr"unglichen Viviane Modell
nur die Bitstrings des modifizierten Viviane-Modells und z"ahlt neben den 
Sprachen {\tt nlang} auch die Sprachfamilien {\tt ifam}. In den zwei 
Zeilen vor dem ersten Print-Befehl muss nach {\tt integer*} bzw.
{\tt data Lg/} die gleiche Zahl 8, 4, 2 oder 1 untereinander angegeben
werden f"ur 64, 32, 16 oder 8 Bits pro Bitstring. Zum Schluss wird ausgedruckt 
die Zahl {\tt ns} der Sprachen einer bestimmten Gr"o"se, die Zahl {\tt nf} 
der Sprachfamilien (mit Gr"o"se = Zahl der Sprachen in der Familie) und das
Histogram {\tt nhist} der Zahl der 1-Bits. Die ersten beiden 
Gr"o"sen werden in Zweierpotenzen zusammengefasst, also z.B. von 32 bis 63.
Weitere Fragen beantwortet bis Anfang 2008 {\tt stauffer@thp.uni-koeln.de}.
  
{\small
\begin{verbatim}
      parameter(L=10000,L2=L*L,L0=1-L,L3=L2+L,L4=25*L+1000,L5=32767,
     1          iscale= 7, imax=11)
c     language colonization of de Oliveira, Gomes and Tsang, Physica A 
c     add bitstring to each language; with Hamming family  analysis
c     ifam(lang) gives the family to which language "lang" belongs
c     nlang(ifam) gives the number of languages within one family
c     grammar bitstring has Lg bytes = 8*Lg bits, integer*Lg grammar
      integer*8 ibm,mult,icount(L5),jpower,numpop,nhist(0:64)
      integer*2 lang,limit,mother(L5),nlang(0:L5)
      byte isite, c
c     byte isite
c     integer*2 c
      dimension neighb(0:3),isite(L0:L3),list(L4),lang(L2),c(L2),f(L5),
     1 nf(0:40),ns(0:40),limit(L5),grammar(0:L5),popct(0:255),ifam(L5)
      integer*8 grammar,grammd,bit(0:63)
      data Lg/8/,iseed/2/,alpha/0.05 /,ns/41*0/,nf/41*0/,nrun/100/
      print *, '# ', L, iseed, alpha, nrun, Lg, iscale, imax, ' >=2'
c     if(iscale.gt.7.or.imax.gt.15) stop 6
      if(Lg.eq.8) kshift=-58
      if(Lg.eq.4) kshift=-59
      if(Lg.eq.2) kshift=-60
      if(Lg.eq.1) kshift=-61
      bit(0)=1
      do 25 i=1,63
 25     bit(i)=ishft(bit(i-1),1)
      mult=13**7
      mult=mult*13**6
      ibm=(2*iseed-1)*mult
      ibm=ibm*16807
      factor=(0.25d0/2147483648.0d0)/2147483648.0d0
      fac=1.0/2.0**iscale
      do 17 j=0,255
        ici=0
        do 18 i=0,7
 18       ici=ici+iand(1,ishft(j,-i))
 17     popct(j)=ici
      neighb(0)= 1
      neighb(1)=-1
      neighb(2)= L
      neighb(3)=-L
      do 11 irun=1,nrun
      call flush(6)
      do 22 ici=0,64
 22     nhist(ici)=0
      numpop=0
      do 10 j=2,L5
        nlang(j)=0
        ifam(j)=0
        mother(j)=0
        icount(j)=0
        ibm=ibm*mult
        limit(j)=1+ishft(ibm,imax-64)
 10     f(j)=0.0
      do 6 j=L0,L3
        if(j.le.0.or.j.gt.L2) goto 6
        lang(j)=0
 9      ibm=ibm*16807
        c(j)=ishft(ibm,iscale-64)
        if(c(j).eq.0) goto 9
        numpop=numpop+c(j)
 6      isite(j)=0
c     print *, limit, ' limit, c ', c
      j=L2/2+1
      isite(j)=1
      isite(j+1)=2
      isite(j-1)=2
      isite(j+L)=2
      isite(j-L)=2
      list(1)=j+1
      list(2)=j-1
      list(3)=j+L
      list(4)=j-L
      isurf=4
      nempty=L2-5
      number=1
      mother(1)=1
      lang(j)=1
      ifam(1)=1
      ifamj=1
      nfam=1
      nlang(nfam)=1
      nlang(0)=0
      grammar(number)=0
      icount(1)=1
      f(1)=c(j)*fac
c     surface=2, occupied=1, empty=0
c     end of initialisation, start of growth
      do 1 itime=1,2000000000
 13     ibm=ibm*16807
        index=1.0+(0.5+factor*ibm)*isurf
        if(index.gt.isurf.or.index.le.0) goto 13
        j=list(index)
c       if(itime.eq.(itime/500000  )*500000  )
c    1   print*,itime,number,isurf,nfam
        ibm=ibm*mult
        if(0.5+factor*ibm .ge. c(j)*fac) goto 1
        list(index)=list(isurf)
        isurf=isurf-1
        isite(j)=1
c       now select language from random neighbour; prob. propto fitness
        fsum=0
        do 5 idir=0,3
 5      if(isite(j+neighb(idir)).eq.1) fsum=fsum+f(lang(j+neighb(idir)))
 3      ibm=ibm*16807
        idir=ishft(ibm,-62)
        i=j+neighb(idir)
        if(isite(i).ne.1) goto 3
        ibm=ibm*mult
        if(0.5+factor*ibm .ge. f(lang(i))/fsum) goto 3
        lang(j)=lang(i)
        ifam(lang(j))=ifam(lang(i))
        grammar(lang(j))=grammar(lang(i))
        f(lang(j))=min(limit(lang(j)), f(lang(j)) + c(j)*fac)
c       now come mutations inversely proportional to fitness f
        ibm=ibm*16807
        if(0.5+factor*ibm .lt. alpha/f(lang(j)) ) then
          number=number+1
          ifamj=ifam(lang(j))
          nlang(ifamj)=nlang(ifamj)+1
          ifam(number)=ifamj
          if(number.gt.L5) stop 8
          ibm=ibm*mult
          nbit=ishft(ibm,kshift)
          grammar(number)=ieor(grammar(lang(j)),bit(nbit))
          lang(j)=number
          f(lang(j))= c(j)*fac
          mother(number)=mother(lang(i))
        end if
        icount(lang(j))=icount(lang(j)) + c(j)
c       now determine Hamming distance (grammd) to previous ancestor
        grammd=ieor(grammar(lang(j)),grammar(mother(lang(i))))
        ici=0
        do 23 ibyte=0,Lg-1
 23       ici=ici+popct(iand(255,ishft(grammd,-8*ibyte)))
        if(ici.ge.2) then
c         new family starts here; subtract previously added language 
          mother(lang(j))=lang(j)
          nlang(ifamj)=nlang(ifamj)-1
          nfam=nfam+1
          ifam(lang(j))=nfam
          if(nfam.ge.L5) stop 7
          nlang(nfam)=1
        end if
        if(isurf.eq.0) goto 8
c       now determine new surface sites as usual in Eden model
        do 2 idir=0,3
          i=j+neighb(idir)
          if(i.le.0.or.i.gt.L2) goto 2
          if(isite(i).ge.1) goto 2
          isurf=isurf+1
          if(isurf.gt.L4) stop 9
          nempty=nempty-1
          list(isurf)=i
          isite(i)=2
 2      continue
 1    continue
 8    continue
      if(L.eq.79) print 7, lang
 7    format(1x,79i1)
      print *, irun, number, itime, numpop, nfam
      do 11 k=1,number
        if(icount(k).gt.0) j=alog(float(icount(k)))/0.69314
        if(k.gt.0.and.k.le.nfam.and.nlang(k).gt.0) then
          i=alog(float(nlang(k)))/0.69314
          nf(i)=nf(i)+1
        end if
 11     ns(j)=ns(j)+1
      jpower=1
      do 12 j=0,37
        if(j.gt.0) jpower=jpower*2
 12     if(ns(j).gt.0) print *, jpower,ns(j),nf(j)
      do 19 j=1,number
        grammd=grammar(j)
        ici=0
        do 20 ibyte=0,Lg-1
 20       ici=ici+popct(iand(255,ishft(grammd,-8*ibyte)))
 19     nhist(ici)=nhist(ici)+1
      do 21 ici=0,64
 21     if(nhist(ici).gt.0) print *, ici, nhist(ici)
      if(nrun.gt.1) stop
      langsum=0
      do 24 i=1,number
        langsum=langsum+nlang(i)
 24     if(nlang(i)+ifam(i).ne.0) print *, i, nlang(i),ifam(i)
      print *, langsum
      stop
      end
\end{verbatim}
}


\end{document}